\documentclass[twocolumn,10pt,pra,aps,showpacs,longbibliography]{revtex4-1}
\usepackage{float}
\floatstyle{boxed}
\usepackage{wrapfig}
\usepackage{array}
\usepackage{graphicx}
\usepackage{amsbsy}
\usepackage{xcolor}
\usepackage[utf8x]{inputenc}
\usepackage{epstopdf}
\usepackage{amsmath,amssymb,amsfonts}
\graphicspath{{pix/}}

\begin{document}

\title{Characterization of  nonclassicality of Gaussian states initially generated in optical spontaneous parametric processes by means of induced stimulated emission}
\author{Ievgen I. Arkhipov}
\email{ievgen.arkhipov@gmail.com}
\affiliation{RCPTM, Joint Laboratory of Optics of Palack\'y University and
Institute of Physics of CAS, Faculty of Science, Palack\'y University, 17. listopadu
12, 771 46 Olomouc, Czech Republic}

\begin{abstract}
In a recent paper [I.~I. Arkhipov, Phys.~Rev.~A {\bf 98}, 021803 (2018)], it was shown that one can completely identify the nonclassicality of  single- and two-mode Gaussian states by means of  certain nonclassicality witnesses which are based on intensity moments up to the third order of optical fields, provided that an appropriate coherent displacement is applied to a given Gaussian state. Here, we utilize a mathematical equivalence between the description of the coherent displaced Gaussian states generated in the spontaneous parametric processes and the Gaussian states generated in the corresponding stimulated parametric processes. Resorting to that equivalence, we study and compare the power of those nonclassicality witnesses in the detection of the nonclassicality of the two-mode Gaussian states generated in both the spontaneous and stimulated second subharmonic  and down-conversion processes and which are subsequently subject to a beam splitter.  We demonstrate that by means of an appropriate induced stimulated emission one can completely  identify the nonclassicality of the considered Gaussian states in comparison to the case of the spontaneous emission. This is important from the experimental point of view, as the stimulated emission can be easily implemented in running optical experiments, and such method can exploit just simple linear detectors.
\end{abstract}

\maketitle

\section{Introduction}
Nonclassical properties of light are at the core of quantum physics, as they allow one to test the most fundamental properties of quantum mechanics. One of the most striking examples of nonclassical behavior of light is entanglement, where spatially separated quanta of light demonstrate strong nonlocal correlations~\cite{Einstein35,Schrodinger35}. The further development of  theory and experiment with entangled light has led to the creation of quantum information theory~\cite{NielsenBook}.

Due to particle-wave dualism of light in quantum mechanics, there are two different classes of quantum states of the optical fields, namely, discrete-variable (DV) and continuous-variable (CV) states, respectively. Mathematically speaking, those two classes of quantum states are distinguished by the size of Hilbert space in which the state vector resides, and  which can be either finite  for DV or infinite for CV states.  

Gaussian states are one of the examples of the CV states. Such states of  light are called Gaussian because the quantum statistical properties of the states can be described by means of the first- and second-order moments of the quantum amplitude (i.e., boson operator) of light. Moreover, Gaussian states are easy to generate in the laboratory and usually they are produced in the optical parametric processes.

The nonclassicality of a two-mode Gaussian state can be expressed by means of two-mode entanglement and local nonclassicality of each mode in terms of squeezing~\cite{AgarwalBook,MandelBook}.

Though the theory of the nonclassicality of the two-mode Gaussian state is well developed, there is still an open question as to how to effectively retrieve that nonclassicality in the experiment.
The state tomography based on homodyne detection  is experimentally demanding; additionally, it requires  further numerical operations for the Wigner function reconstruction (e.g., Radon transformations or maximum-likelihood algorithm~\cite{Banaszek1999,Rehacek2007}), the negative values of which characterize the state nonclassicality~\cite{LeonhardtBook,Lvovsky2009}. Moreover, in the case of Gaussian states, the Wigner function is always positive~\cite{AgarwalBook} and as such one needs to resort rather to the reconstructed Glauber-Sudarshan $P$ function~\cite{Glauber1963,Sudarshan1963}, the negativity or nonregularity of which explicitly points to the nonclassicality of the state.  On the other hand, one can retrieve the Wigner function of the measured  optical field without the need for the homodyne tomography, but directly from photocount statistics provided that the appropriate coherent displacement of the state takes place~\cite{Banaszek1999,Bondani2010,Pfister2014,Kuhn2016}. 

 Other approaches relying on the direct measurements of photon numbers and integrated  intensity moments  have been proposed~\cite{Agarwal92,Klyshko1996,Achilles2003,Fitch2003,Haderka2004,Avenhaus2010,Sperling2015,Blanchet2008,Lee1990a,Shchukin05,Miranowicz2010,Sperling2015,PerinaJr2017a,arkhipov2018a}.

All of the above proposed methods are based on measuring the so-called nonclassicality witnesses (NWs), e.g., some classical inequalities containing fields moments that when violated (e.g., become negative), they identify the presence of the nonclassical correlations in the measured state. Some of such NWs cannot  guarantee reliable extraction of the nonclassicality of the state if one applies them directly to the measured state, since some form of the nonclassicality might not be accessible by NWs, and therefore additional transformations in the phase space of the measured state are needed to reach the hidden nonclassicality by the given NWs~\cite{arkhipov2018a}. 

In a recent paper~\cite{arkhipov2018c}, a method which enables one to completely reveal the nonclassicality of any single- and two-mode  Gaussian state by means of integrated intensity moments up to the third order was proposed. That approach again relies on a specific form of NW but requires  certain coherent displacements of the state under consideration. 

The Gaussian states generated in the stimulated parametric processes are mathematically equivalent to the coherent displaced Gaussian states initially generated in the spontaneous parametric processes~\cite{Perina1991Book}. Moreover, the nonclassical properties of the Gaussian states with coherent components are independent of those coherent terms, meaning that stimulated emission does not affect the nonclassicality properties of the Gaussian states initially generated in the corresponding spontaneous parametric processes.

Motivated by that equivalence and the results in Ref.~\cite{arkhipov2018c},  we present a study of the complete experimental identification of the nonclassicality of Gaussian states generated in stimulated parametric processes using the example of the stimulated second  sub-harmonic and down-conversion processes. Additionally, we compare the power of the considered NWs with the genuine nonclassicality identifiers for local squeezing of each mode and entanglement between modes of the Gaussian state~\cite{arkhipov2016b,arkhipov2016c}. 
As such, we demonstrate that by means of the induced stimulated emission, provided that an appropriate phase of the stimulating coherent field is chosen, one can completely  identify the nonclassicality of the Gaussian states in comparison to the case when the similar Gaussian states are generated in the process of the spontaneous emission. Furthermore, in comparison to the aforementioned methods which are based on NW, our approach guarantees the full extraction of the nonclassicality of the Gaussian states in the whole phase space, and which does not require either a balanced or an unbalanced homodyne detection. Hence we present a reliable, simple, and convenient technique for identification of the nonclassicality of Gaussian states of light generated in the parametric processes and which subsequently undergo beam splitter transformations. Despite the fact that in the presented study we focus mainly on single - and two-mode Gaussian states generated in the considered parametric processes, our method can be easily extended to the multimode scenario.


We note that the study of the nonclassical properties of the stimulated twin beam in terms of the quasiprobability function of integrated intensities has been given in Ref.~\cite{Perina2005}. 
 Moreover, the stimulated parametric processes have found a wide range of application in the generation and characterization of various nonclassical states of light~\cite{Resch2002,Zavatta2004,Zavatta2007,Keshari2013}.

The paper is organized as follows. In Sec.~II, we introduce the theory of the two-mode Gaussian states generated in both the spontaneous and stimulated second subharmonic and down-conversion parametric processes, and  which is subsequently subjected to the beam splitter transformations. Additionally, we present the nonclassicality witnesses which are at the core of the given study.  In Sec.~III, we apply those NWs to characterize and compare their power in the identification of the nonclassicality of single-mode Gaussian states produced in both the spontaneous and stimulated subharmonic generation process with successive beam splitter transformation. We give the same analysis, but for the case of the two-mode Gaussian state generated in the spontaneous and stimulated down-conversion processes, in Sec.~IV. In Sec.~V, we briefly discuss the generalization of the obtained results in the previous sections for a multimode case. Conclusions are drawn in Sec.~V.

\section{Theory}
\subsection{Spontaneous and stimulated second subharmonic and down-conversion processes}
In what follows, we investigate the parametric subharmonic (SH) and parametric frequency down-conversion (DC) processes during spontaneous and stimulated emissions.
To describe the Gaussian states generated in the stimulated optical parametric processes, it is easier to start with the consideration of the spontaneous processes since the stimulation effect can be added straightforwardly.

The interaction Hamiltonian describing both spontaneous SH and DC processes for a two-mode field with ideal phase-matching conditions in the parametric approximation can be written in the following form:
\begin{equation}\label{H}  
\hat H_{int}= -\hbar g_{1}^*\hat a_1^2-\hbar g_{2}^*\hat a_2^2 -\hbar g_{3}^*\hat a_1\hat a_2 + {\rm h.c.},
\end{equation}
where $g_l$, $l=1,2,3$, are, in general, complex coupling constants, $\hat a_k$ is the boson annihilation operator of the mode $k$, $k=1,2$, and H.c. stands for the Hermitian conjugated terms.
The parametric approximation is based on the assumption that the pump fields generating SH and DC quantum fields are in coherent states and thus are treated as classical waves, i.e., their amplitudes are incorporated in the scalar couplings $g_l$ in Eq.~(\ref{H}). The first two terms in Eq.~(\ref{H}) describe the second subharmonic generation process of modes 1 and 2, and the third term accounts for the down-conversion process.

The real nonlinear processes always include some amount of noise in the system, which is represented as a reservoir of the 'noisy' optical phonons interacting with the optical field in mode $k$ and have the mean phonon number  $\langle n_{dk}\rangle=\left[\exp\left(\hbar\psi_k/KT\right)-1\right]^{-1}$, where $\psi_k$ is the frequency of the reservoir mode $k$, $T$ and $K$ are the temperature and the Boltzmann constant, respectively, and the damping coefficient $\gamma_k$ stands for damping of the $k$th propagating optical mode.

To mathematically incorporate the mentioned noise for the system described by the Hamiltonian in Eq.~(\ref{H}), one resorts to the Heisenberg-Langevin operator equations which take the following form:
\begin{eqnarray} \label{hle}    
 \frac{d\boldsymbol{\hat a}}{dt} &=& \boldsymbol{M}\boldsymbol{\hat a} + {\boldsymbol{
\hat  L}}
\end{eqnarray}
where $\boldsymbol{\hat a} = (\hat a_1,\hat a_1^{\dagger},\hat a_2,\hat a_2^{\dagger})^T$, and  $\boldsymbol{
\hat  L} = \left(\hat L_1,\hat L_1^{\dagger},\hat L_2,\hat
L_2^{\dagger}\right)^T$, and the operators of the Langevin fluctuating forces $\hat L_1 $ and $\hat L_2 $ obey the following relations:
\begin{eqnarray}            
\langle\hat L_{i}(t)\rangle &=& \langle\hat
L^{\dagger}_{i}(t)\rangle= 0, \nonumber \\
 \langle\hat L^{\dagger}_{i}(t)\hat L_{j}(t')\rangle &=&
\delta_{ij}\gamma_i\langle n_{di}\rangle\delta(t-t'), \nonumber \\
\langle\hat L_{i}(t)\hat L^{\dagger}_{j}(t')\rangle
&=&\delta_{ij}\gamma_i\big( \langle n_{di}\rangle +1\big)\delta(t-t'),
\end{eqnarray}
where $\delta_{ij}$ stands for the Kronecker symbol and $\delta$ denotes
the Dirac delta function.

The matrix $\boldsymbol{M}$ in Eq.~(\ref{hle}) for Hamiltonian $H_{int}$ in Eq.~(\ref{H}) acquires the form
\begin{equation}
\boldsymbol{M}=\begin{pmatrix}
0 & 2ig_1 & 0 & ig_3 \\
-2ig_1^* & 0 & -ig_3^* & 0 \\
0&  ig_3 & 0 & 2ig_2 \\
-ig_3 & 0 & -2ig_2 & 0
\end{pmatrix}.
\end{equation}
For time-independent matrix $\boldsymbol{M}$ the solution for the operators $\boldsymbol{\hat a}(t)$ in Eq.~(\ref{hle}) can be written as
\begin{equation}
\boldsymbol{\hat a}(t)=\exp(\boldsymbol{M}t)\boldsymbol{\hat a}(0)+\int\limits_0^t{\rm d}t'\exp\left(\boldsymbol{M}(t-t')\right)\boldsymbol{\hat L}(t').
\end{equation}
The elements of the operators $\boldsymbol{\hat a}(t)$  can be expressed as
\begin{equation}\label{UV}
\hat a_j(t) = \sum\limits_{l=1}^2\left[U_{jl}(t)\hat a_l(0)+V_{jl}(t)\hat a_l^{\dagger}(0)\right] +\hat F_j,
\end{equation}
where the matrices $\boldsymbol{U}$, $\boldsymbol{V}$, and $\boldsymbol{\hat F}$ are obtained from the eigenvalues and the matrix of the eigenvectors of  $\boldsymbol{M}$ (for details, see Ref.~\cite{PerinaJr2000}). 

Now when considering the stimulation parametric processes, we need to take into account the dynamics of the coherent fields which provide stimulation. Regarding the evolution of the coherent field $\xi_j(t)$, $j=1,2$, which stimulate the SH and DC emission of the $j$th mode, one can write the  expressions~\cite{PerinaJr2000}
\begin{equation}\label{CUV}
\xi_j(t)=\sum\limits_{l=1}^2U_{jl}(t)\xi_l(0)+V_{jl}(t)\xi_l^{*}(0),
\end{equation} 
where $\xi_j(0)$ accounts for the amplitude of the coherent field which stimulates the mode $j$ at time $t=0$.

All the quantum statistical information of the two-mode Gaussian state described by the Hamiltonian in Eq.~(\ref{H}) can be described by the corresponding normal characteristic function  written as
\begin{equation}\label{CN}	 
C_{\cal N}(\beta_1,\beta_2)=\exp\left(-\frac{1}{2}\boldsymbol{\beta}^{\dagger}\boldsymbol{\Omega}\boldsymbol{A}_{\cal N}\boldsymbol{\Omega}^T\boldsymbol{\beta}+\boldsymbol{\beta}^{\dagger}\boldsymbol{\Omega}\boldsymbol{\Xi}\right),
\end{equation}
where $\boldsymbol{\beta}=(\beta_1^*,\beta_2,\beta_2^*,\beta_2)^T\in{\mathbb C}^4$ is a complex vector,  $[\boldsymbol{A}_{\cal N}]_{jk}=\langle :\!\!\Delta\hat A_j^{\dagger} \Delta\hat A_k\!\!:\rangle =\langle :\!\! \hat A_j^{\dagger} \hat A_k\!\!:\rangle - \langle \hat A_j^{\dagger}\rangle\langle \hat A_k\rangle$ are the elements of the normal covariance matrix with $\hat A = \left[\hat a_1^{\dagger}(t),\hat a_1(t),\hat a_2^{\dagger}(t),\hat a_2(t)\right]^T$,  and where the matrix  $\boldsymbol{\Omega}=\bigotimes\limits_{k=1}^{2}\omega_k$, and $\omega_k=\begin{pmatrix}0&1 \\ -1 & 0\end{pmatrix}$.  And one has the complex vector $\boldsymbol{\Xi}=\left[\xi_1(t),\xi_1^*(t),\xi_2(t),\xi_2^*(t)\right]^T$.

The covariance matrix $\boldsymbol{ A_{\cal N}}$ of the two-mode field in Eq.~(\ref{CN})
can be written explicitly in the following block form:
\begin{eqnarray}\label{AN}    
 \boldsymbol{ A_{\cal N}} &=& \left( \begin{array}{cc} {\bf B}_1 & {\bf
  D}_{12}  \cr {\bf D}_{12}^{\dagger} & {\bf B}_2 \end{array} \right) ,
\label{11} \\
 {\bf B}_j &=& \left( \begin{array}{cc} B_j & C_j
  \cr C_j^* & B_j \end{array} \right) , \hspace{3mm} j=1,2,
  \nonumber \\
 {\bf D}_{12} &=& \left( \begin{array}{cc} \bar{D}_{12}^* & {D}_{12}
  \cr {D}_{12}^* & \bar{D}_{12} \end{array} \right),
\end{eqnarray}
where the block matrices ${\bf B}_1$ and ${\bf B}_2$ are responsible for local quantum correlations of the corresponding reduced modes. The block matrix ${\bf D}_{12}$ describes the quantum cross-correlations between modes.

It is important to note that all of the nonclassical features of the Gaussian states are completely encoded into the corresponding covariance matrix, and thus the coherent part of the quantum state plays no role in the generation of the nonclassicality, as it can always be displaced to vacuum. Nevertheless, as it has been shown in Ref.~\cite{arkhipov2018c}, and which  we will show here later, the coherent part of the Gaussian state has  huge importance in the problem of the experimental extraction of the nonclassicality of the state. 

\subsection{Beam splitter transformation}
Since in the following sections we will focus on the single- and two-mode Gaussian states generated by the Hamiltonian in Eq.~(\ref{H}) and which subsequently undergo the beam splitter transformations, we would also like to consider the very action of such transformation on the considered states.

In general, a beam splitter action on a two-mode Gaussian state can be expressed by the corresponding unitary transformation $\bf{S}$ which has the following matrix form:
\begin{eqnarray}\label{BS}       
 &&{\bf S}=\left(\begin{array}{cccc}
  \sqrt{T} & 0 &-\sqrt{R}e^{i\theta}&0 \\
  0 & \sqrt{T}& 0 & -\sqrt{R}e^{-i\theta}  \\
  \sqrt{R}e^{-i\theta} & 0 &\sqrt{T} & 0\\
  0 & \sqrt{R}e^{i\theta} & 0 & \sqrt{T}\end{array}\right),
\end{eqnarray}
where $T$ is a transmissivity of the beam splitter, and $R=1-T$. The phase $\theta$ occurring in Eq.~(\ref{BS}) can be set
to zero without the loss of generality. In what follows, we always assume that $T$ stands for transmissivity.

The beam splitter with the unitary matrix $\bf{S}$ in Eq.~(\ref{BS})  just transforms the corresponding covariance matrix $\boldsymbol{ A_{\cal N}}$ and the complex vector $\bf{\Xi}$ which determine the characteristic function $C_{\cal N}$ as follows:
\begin{equation}\label{BSaction}
\boldsymbol{ A_{\cal N}}\rightarrow{\boldsymbol{S^{\dagger}}}\boldsymbol{ A_{\cal N}}{\boldsymbol{S}}, \quad \boldsymbol{\Xi}\rightarrow\boldsymbol{S^{\dagger}}\boldsymbol{\Xi}.
\end{equation}


\subsection{Nonclassicality criteria of Gaussian states based on the integrated intensity moments}
As has been found in Ref.~\cite{arkhipov2018c}, the most powerful nonclassicality witnesses (NWs) based on the integrated intensity moments for single-mode and two-mode Gaussian fields  $ \langle W_1^mW_2^n\rangle$ can be written as
\begin{eqnarray}\label{NW}  
R_k&=& \langle W_k\rangle\langle W_k^3\rangle - \langle W_k^2\rangle^2<0, \quad k=1,2 \\
M &=& \langle W_1^{2}\rangle \langle W_2^2\rangle - \langle W_1W_2\rangle^2<0.
\end{eqnarray}

The NWs in Eq.~(\ref{NW}) can completely reveal the nonclassicality either local or between two modes provided that an appropriate coherent displacement is applied to the state~\cite{arkhipov2018c}.

We would like to stress, that the NWs in Eq.~(\ref{NW}) can, in general, be used for the nonclassicality detection of any kind of state of light, e.g., including non-Gaussian states, since these NWs, when negative, point to the nonclassicality of the Glauber-Sudarshan $P$ function, which plays a role of the very definition of the nonclassicality of light~\cite{AgarwalBook,Shchukin2006,PerinaJr2017a}.
  
The integrated intensity moments  can be found via the normal generating function $G_{\cal N}$, which for the two-mode state  is defined as
\begin{eqnarray}   
G_{\cal N}(\lambda_1,\lambda_2)=\frac{\exp\left(-\frac{1}{2}\boldsymbol{\Xi^{\dagger}}\boldsymbol{{ A}'_{\cal N}}^{-1}\boldsymbol{\Xi}\right)}{\lambda_1\lambda_2\sqrt{{\rm det}\boldsymbol{ A'_{\cal N}}}},
\end{eqnarray}
with $\boldsymbol{ A'_{\cal N}}=\boldsymbol{ A_{\cal N}}+\boldsymbol{\lambda}^{-1}\mathbb{I}_4$, where $\mathbb{I}_4$ is a four dimensional identity matrix, and we denote the matrix $\boldsymbol{\lambda}^{-1}={\rm diag}(1/\lambda_1,1/\lambda_1,1/\lambda_2,1/\lambda_2)$.

The integrated intensity moments $ \langle W_1^{k_1}W_2^{k_2}\rangle
$ are obtained along the formula
\begin{eqnarray}  
\hspace{-3mm} \langle W_1^{k_1}W_2^{k_2}\rangle &&= (-1)^{k_1+k_2} \nonumber \\
&&\times\left.\frac{\partial^{k_1+k_2}  G_{\cal   N}(\lambda_1,\lambda_2)}{\partial\lambda_1^{k_1}\partial\lambda_2^{k_2}}\right|_{\lambda_1=\lambda_2=0}.
\end{eqnarray}

Here, we would like to stress that when analyzing below NWs $R_k$ and $M$ for the Gaussian states under consideration, we will compare those NWs with genuine nonclassicality identifiers for the two-mode Gaussian states derived in Refs.~\cite{arkhipov2016b,arkhipov2016c}. Namely, those genuine nonclassicality identifiers comprise the local nonclassicality identifiers (LNIs) $I_{\rm ncl}^{(1)}$ and $I_{\rm ncl}^{(2)}$, which fully characterize the local nonclassicality of  modes 1 and 2, respectively, and the entanglement identifier (EI) $I_{\rm ent}$, which accounts for the entanglement between two modes. Most importantly, these LNIs and EI form the nonclassicality invariant which holds true for any unitary photon-number preserving operations, namely~\cite{arkhipov2016b},
\begin{equation}\label{GNI}  
I_{\rm ncl}=I_{\rm ncl}^{(1)}+I_{\rm ncl}^{(2)}+2I_{\rm ent},
\end{equation}
that allows us unmistakenly to evaluate the power of the NWs given in Eq.~(\ref{NW}) with respect to the present genuine local and inter-mode nonclassicality of the quantum state.

We would like to note that a similar nonclassicality invariant has been recently found and experimentally verified for discrete variable states, i.e., for qubits~\cite{Svozilik15,Cernoch2018}.

\section{revealing nonclasicality of Gaussian states generated in second subharmonic processes by means of stimulated emission}
\subsection{Spontaneous Second Subharmonic Emission}
For the pure single-mode Gaussian state generated in the spontaneous second subharmonic process, i.e., only coupling $g_1$ is presented in the Hamiltonian (\ref{H}),  the nonzero elements of the matrices $\bf U$ and $\bf V$ in Eq.~(\ref{UV}) have the  form
\begin{equation}\label{spUVsq}
U_{11}=\cosh(2g_1t), \quad V_{11}=ie^{i\alpha}\sinh(2g_1t),
\end{equation}
where $\alpha$ is the phase of the pumping field, which, without loss of generality, we can take to be $\alpha=0$, and, for simplicity, we assume that the coupling constant $g_1$ is a real-valued number.

The Gaussian states generated in such spontaneous second subharmonic generation processes are called squeezed vacuum states, as they reveal the squeezing of one of its quadratures~\cite{MandelBook,AgarwalBook}.

 The nonzero elements  of the normal covariance matrix $\boldsymbol{ A_{\cal N}}$ 
are the mean photon number $B_1$, which describes the mean photon number of the squeezed vacuum, i.e.,  $B_1=B_{ \rm sq}$, where $B_{\rm sq}= |V_{11}|^2$, and the complex parameter $C_1=i\sqrt{B_{\rm sq}(B_{\rm sq}+1)}$, which account for the squeezing effect.

To include noise in the system, we utilize the model of the superposition of the quantum signal and noise~\cite{Perina1991Book}, meaning that only the vacuum fluctuations $B_1$ of the signal field need to be modified in the covariance matrix, i.e., $B_1=B_{\rm sq}+B_{\rm n}$,  where $B_{\rm n}$ is the mean thermal noise photon number, and all other quantities in the covariance matrix $\boldsymbol{ A_{\cal N}}$ are left unchanged~\cite{arkhipov2016c}.
\begin{figure} 
\includegraphics[width=0.33\textwidth]{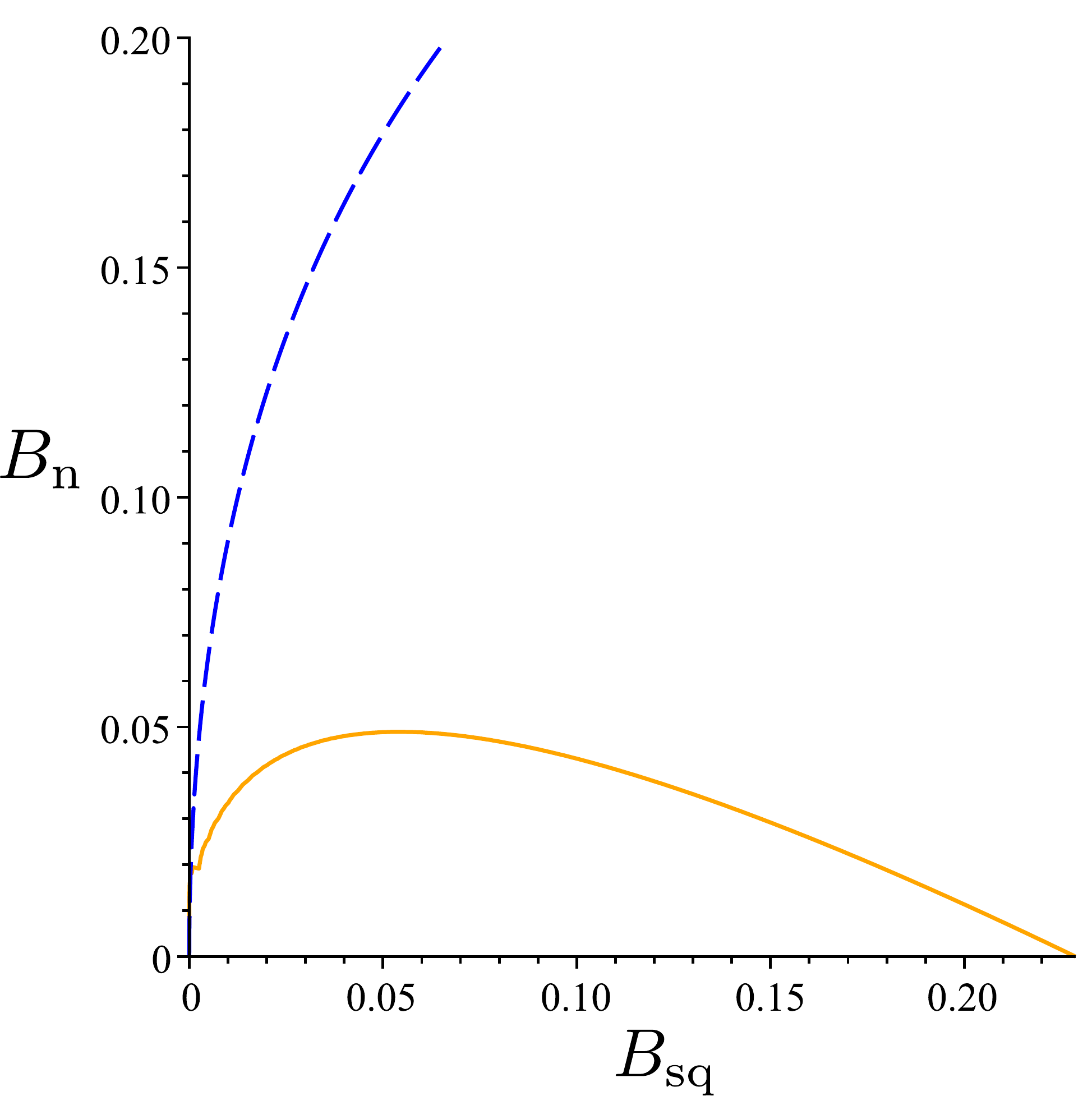}
\caption{Phase diagram of negative values of function $f$ giving NWs $R_1$ and $R_2$ in Eq.~(\ref{19}) (orange solid curve), for a squeezed vacuum state in space spanned by a mean squeezed photon number $B_{\rm sq}$ and mean noise photon number $B_{\rm n}$. The region of negativity of $f$ is  below the curve. For comparison, the nonclassicality phase diagram of LNI $I_{\rm ncl}^{(1)}=I_{\rm ncl}^{(2)}$ in Eq.~(\ref{GNI}) (blue dashed curve) is shown on the graph, and the region below this curve identifies the region of  squeezed nonclassical states.}\label{fig1}
\end{figure}

By mixing the squeezed vacuum mode $\hat a_1(t)$ [with the solution given in Eq.~(\ref{UV}) and provided Eq.~(\ref{spUVsq})] with vacuum on the beam splitter, one arrives at the variety of states which are both locally squeezed and entangled at the output of the beam splitter depending on the transmissivity $T$. Such output states at the beam splitter have been extensively studied in Ref.~\cite{arkhipov2016c}.  

Now, we would like to study the power of the NW $R_k$ for mode $k=1,2$ at the output of the beam splitter in the revealing of the local nonclassicality of the output state. Hence we apply the NWs $R_1$ and $R_2$ to the output modes 1 and 2 of the BS, respectively,  to quantify each mode's local squeezing. 
The NWs $R_1$ and $R_2$ can be written as 
\begin{equation}\label{19}
R_1 = f(B_{\rm sq},B_{\rm n})T^4, \quad R_2 = f(B_{\rm sq},B_{\rm n})(1-T)^4.
\end{equation}
 Thus, the negativity of NW $R_k$ does not depend on the transmissivity $T$, but on the negativity of the function $f$. The behavior of the function $f$ along with genuine local nonclassicality identifiers  $I_{\rm ncl}^{(1)}=I_{\rm ncl}^{(2)}$ are shown in Fig.~\ref{fig1}.
As it can be seen, the function $f$ attains negative values only for a small range of the mean photon-number $B_{\rm sq}\in(0,1/4[\sqrt{11/3}-1])$. Moreover, the sensitivity of the negativity of $f$ to the noise is very high; as a result, the NW $R_k$, $k=1,2$, already becomes positive for a quite small amount of noise. 

As a result, the NW $R_k$, $k=1,2$, fails, in general, to identify the local nonclassicality of the quantum state generated in the spontaneous second subharmonic process and which is subsequently subject to the beam splitter with arbitrary transmissivity $T$. Additionally, it turns out that the NW $M$ which can account for the entanglement of the state at the output of BS  is everywhere positive for any $B_{\rm sq}$ and $T$ even for the pure state. 

\subsection{Stimulated Subharmonic Emission}
Now, we would like to see what happens with the power of NW $R_k$  if one applies the stimulating coherent field $\xi_1(0)=|\xi_1|e^{i\phi_1}$ to the subharmonic process of the generated single-mode 1. 
\begin{figure} 
\includegraphics[width=0.35\textwidth]{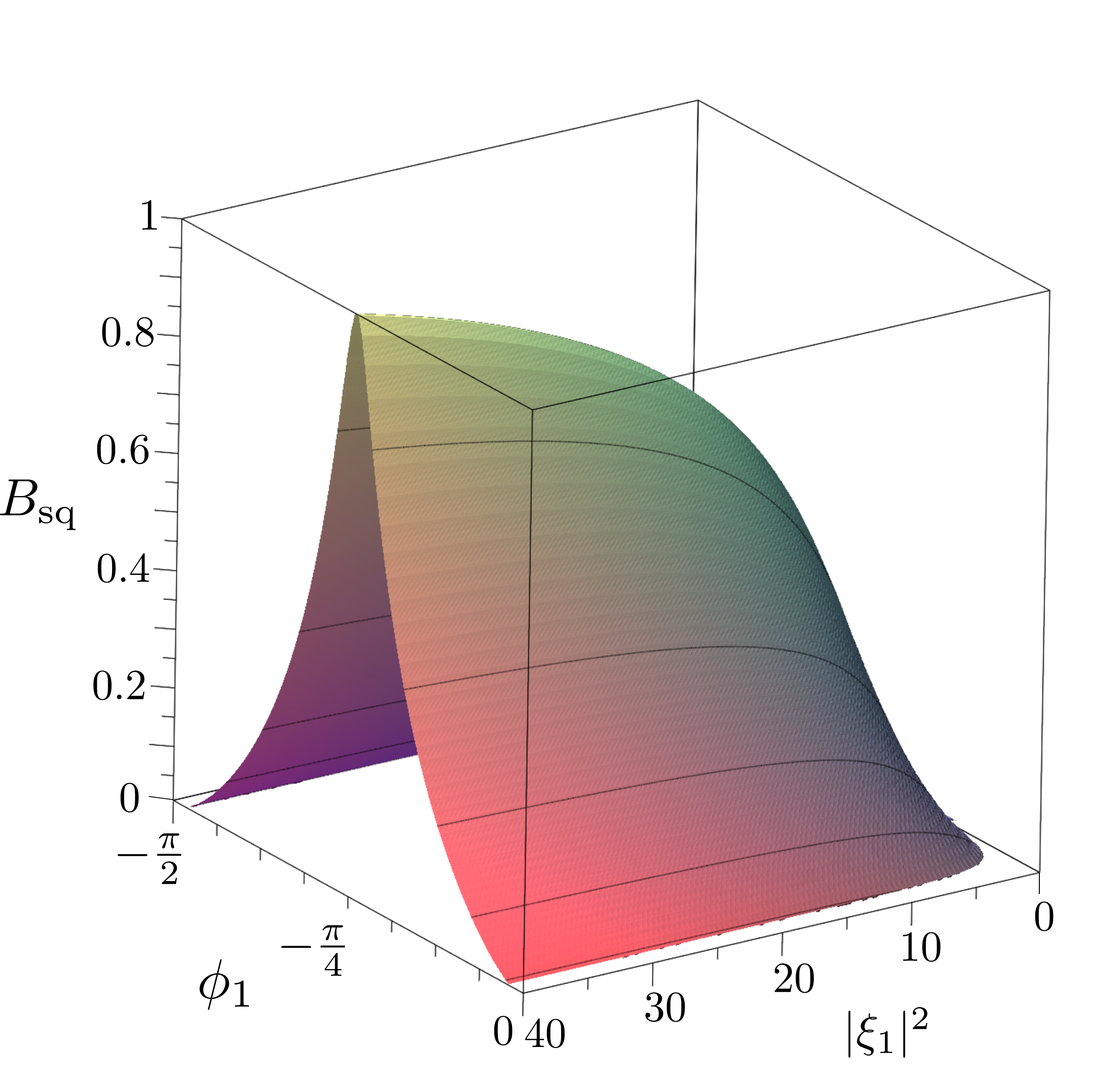}
\caption{Phase diagram of negative values of  $f$  giving NWs $R_1$ and $R_2$ in Eq.~(\ref{20}) for a stimulated squeezed vacuum state in space spanned by intensity $|\xi_1|^2$ and the phase $\phi_1$  of the stimulating coherent field, and the vacuum fluctuations $B_{\rm sq}$, assuming $B_{\rm sq}=1$ and $B_{\rm n}=0$. The region inside the $\Lambda$-shaped volume identifies the region of negativity of $f$. } \label{fig2}
\end{figure}

\begin{figure} 
\includegraphics[width=0.35\textwidth]{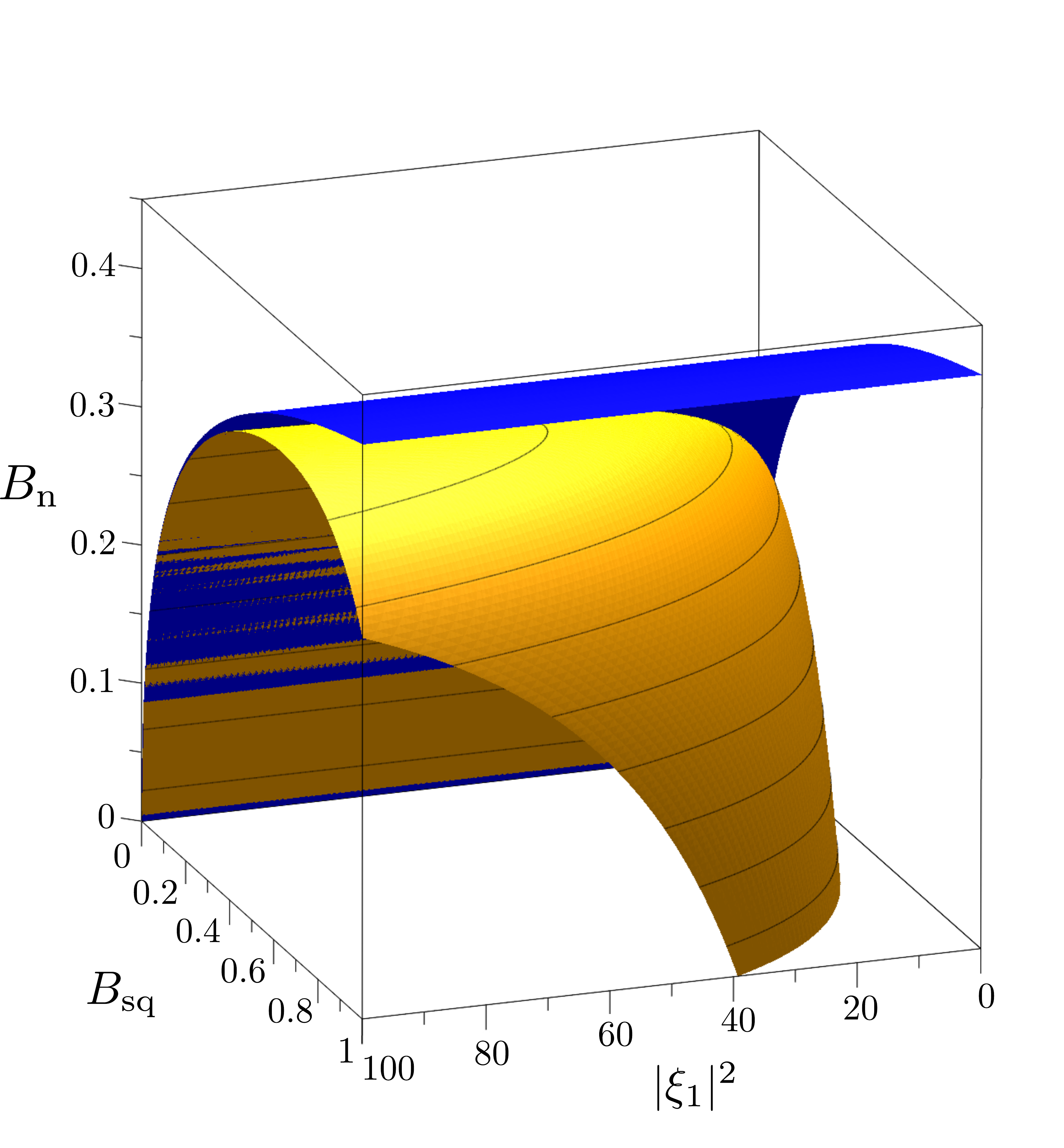}
\caption{Phase diagram of negative values of function $f$ giving NWs $R_1$ and $R_2$ in Eq.~(\ref{20}) (orange  surface)  for a stimulated noisy squeezed vacuum state in space spanned by mean squeezed photon number $B_{\rm sq}$, mean noise photon number $B_{\rm n}$, and real amplitude of the stimulating coherent intensity $|\xi_1|^2$.  The regions under the surface denote the negative values of $f$. For comparison, the nonclassicality phase diagram of LNI $I_{\rm ncl}^{(1)}=I_{\rm ncl}^{(2)}$ (blue surface) is shown on the graph.}\label{fig3}
\end{figure}
We recall that in the case of the stimulated parametric emission, the characteristic function $C_{\cal N}$ in Eq.~(\ref{CN}) acquires the coherent term described by the vector $\bf \Xi$, and, in the case of the stimulated subharmonic emission, the coherent stimulating field $\xi_1(t)$ takes the form according to  Eq.~(\ref{CUV}) for the given $\bf U$ and $\bf V$ in Eq.~(\ref{spUVsq}).

The mean photon number $\langle n_{\rm st}\rangle$ of the state generated in the stimulated  subharmonic emission  can be  written as $\langle n_{\rm st}\rangle = B_1 + |\xi_1(t)|^2$. As such the  $\langle n_{\rm st}\rangle$ is decomposed into the quantum part $B_1=B_{\rm sq}+B_{\rm n}$,  where $B_{\rm sq}$ accounts for the vacuum fluctuations of the signal and which is related to the mean squeezed photon number in the case of the spontaneous emission, but now it corresponds just to the vacuum fluctuations of the stimulated field,  $B_{\rm n}$ is the mean photon number of the thermal noise, and   $|\xi_1(t)|^2$ is the coherent intensity of the stimulating field.

The NWs $R_1$ and $R_2$ are just transformed as 
\begin{eqnarray}\label{20}
R_1 &=& f(B_{\rm sq},B_{\rm n},\xi_1(t))T^4, \nonumber \\
R_2 &=& f(B_{\rm sq},B_{\rm n},\xi_1(t))(1-T)^4,
\end{eqnarray}
 meaning that they preserve their factorized form between transmissivity $T$  and the rest parameters, as in the case of the spontaneous emission. 

Remarkably, but in the case of the stimulated second subharmonic emission, one can completely retrieve the local nonclassicality presented in the system with NW $R_k$ by choosing appropriate phase $\phi_1$ and intensity $|\xi_1|^2$ of the stimulating coherent field  $\xi_1(0)=|\xi_1|e^{i\phi_1}$. For the pure state,  one can detect the local  nonclassicality only within a certain range of values of the phase $\phi_1$ centered at the point $\phi_1=-\pi/4+\pi n$, $n\in \mathbb{Z}$, as shown in  Fig.~\ref{fig2}. Also, as  Fig.~\ref{fig2} infers, the larger the values of $B_{\rm sq}$, the smaller the range of the $\phi_1$ in which one can access the nonclassicality, meaning that in the limit $B_{\rm sq}\rightarrow\infty$, the only possible phase  is $-\pi/4+\pi n$.

The above result, concerning the optimal phase $\phi_1$ needed to reveal the local nonclassicality, can be obtained analytically. Indeed, as  shown in Ref.~\cite{arkhipov2018c}, the optimal phase of the coherent field $\xi_1(t)$, required for the negativity of the NW $R_k$ for the squeezed state, is exactly $-\pi/4+\pi n$. But that value of the phase is for the modulated coherent field $\xi_1(t)$ which is at the output of the beam splitter not of initial $\xi_1(0)$. By making a simple calculation, one immediately arrives to the optimal phase $\phi_1$ of the initial stimulating field $\xi_1(0)$ which also equals the value $-\pi/4+\pi n$.

By adjusting the initial phase $\phi_1$ of the coherent field $\xi_1(0)$ to the value $-\pi/4+\pi n$, one is able to identify the local nonclassicality for any values of the vacuum fluctuations $B_{\rm sq}$ by choosing the appropriate critical value of the initial coherent intensity $|\xi_1|^2$ (see Fig.~\ref{fig3}). As  Fig.~\ref{fig3} certifies, the larger the values of the $B_{\rm sq}$, the larger the values of $|\xi_1|^2$ are needed to make NW $R_k$ negative. The same analysis applies to the existing noise $B_{\rm n}$ in the system; namely, for larger noisy beams, the larger coherent intensities are required to identify the nonclassicality. Moreover, the local nonclassicality of the state which is actually generated by the vacuum fluctuations $B_{\rm sq}$ can be extracted by NW $R_k$ for any amount of noise which does not conceal the nonclassicality, though one needs to resort to the greater values of the coherent intensities (see Fig.~\ref{fig3}).

Regarding the NW $M$,  it displays positivity everywhere for any $\xi_1(t)$. In other words, one cannot rely on NW $M$ to detect two-mode entanglement at the output of the BS even in the presence of the stimulated emission, though the NW $M$, as shown in Ref.~\cite{arkhipov2018c}, can also characterize local nonclassicalities. Nevertheless, one can still identify the amount of the entanglement by means of the nonclassicality invariant $I_{\rm ncl}$ in Eq.~(\ref{GNI}).  The LNIs $I_{\rm ncl}^{(1)}$ and $I_{\rm ncl}^{(1)}$ are proportional to the NWs $R_1$ and $R_2$, respectively, namely, as $I_{\rm ncl}^{(k)}=\chi R_k$, $k=1,2$, where $\chi=\chi(B_{\rm sq},B_{\rm n},|\xi_1|)$. Note that the signs of $\chi$ and $R$ are opposite, as negative values of $R_k$ always lead to the positive LNI $I_{\rm ncl}^{(k)}$, and vice versa. 
As the global invariant $I_{\rm ncl}$ in Eq.~(\ref{GNI}) suggests, the initial nonclassicality of the single-mode stimulated second subharmonic field is solely expressed by the LNI $I_{\rm ncl}^{(1)}[1]$, where in the square parentheses we denote the value of the transmissivity of the beam splitter.  After the beam splitter with arbitrary $T$, the global nonclassicality invariant consists of the two LNIs   $I_{\rm ncl}^{(1)}[T]$ and $I_{\rm ncl}^{(1)}[T]$ and the entanglement identifier $I_{\rm ent}[T]$. By putting all this together, one obtains
\begin{equation}  
I_{\rm ncl}^{(1)}[1]=I_{\rm ncl}^{(1)}[T]+I_{\rm ncl}^{(1)}[T]+2I_{\rm ent}[T].
\end{equation}
From the latter, it follows that the entanglement can be observed whenever
\begin{equation} 
I_{\rm ent}[T]\equiv I_{\rm ncl}^{(1)}[1]-I_{\rm ncl}^{(1)}[T]-I_{\rm ncl}^{(1)}[T]>0.
\end{equation}
By expressing now the LNIs by means of NWs $R_1[T]$ and $R_2[T]$, and taking into account the assumed negativity $R_k[T]$, one arrives finally to
\begin{equation}\label{23}  
I_{\rm ent}[T]\equiv R_1[T]+R_2[T]-R_1[1]>0.
\end{equation}
The derivation of Eq.~(\ref{23}) requires that the total photon number of the whole system is preserved~\cite{arkhipov2016b}.


\section{Revealing nonclassicality of a twin beam on BS by means of stimulated twin beam emission}
\subsection{Twin Beam on BS}
The study of the power of the NWs $R_k$, $k=1,2$, and $M$ for the twin beam generated in the spontaneous parametric down-conversion process, which are subsequently subject to the BS, has been provided in Ref.~\cite{arkhipov2018a}. Here, we just briefly summarize the main results which have been obtained in that study. First,  it has been shown that the NW $M$ serves as a genuine NW only for a free-propagating noisy twin beam, i.e., with BS transmissivity $T=1,0$. Also for $T\neq1,0$, the power of NW $M$ quickly drops as $T\rightarrow1/2$, as such the NW $M$ in general fails to identify the nonclassicality of the twin beam with or without noise for arbitrary $T\neq1,0$. The same conclusions have been drawn for the NWs $R_1$ and $R_2$ which demonstrate a good local nonclassical sensitivity only in the case of a pure twin beam on BS. For noisy twin beams on BS, the NWs $R_k$, $k=1,2$, exhibit lower power in the identification of the local squeezing of each mode. From the presented results, it could be inferred that in general, one cannot rely on the considered NWs in the  problem of the identification of the nonclassicality of, in general, a noisy twin beam state passing through the BS with $T\in(0,1)$.

\subsection{Stimulated Twin Beam on BS}
In the previous section, it was shown that the identification of the nonclassicality  of the Gaussian states obtained in the  stimulated emission of the second subharmonic process by means of the NWs $R_k$, $k=1,2$ can lead to their drastic improvement in the nonclassicality detection; the same holds true for the NWs $R_k$ and $M$ applied to the twin beam states generated in the stimulated down-conversion process. This means that the extraction of the nonclassicality of the twin beam on BS  can be fully performed by means of the considered NWs, provided that the appropriate stimulating coherent fields are applied to the twin beam state.

In what follows, we pay attention to the twin beam which is stimulated only in one mode, let us say in the signal mode, i.e., $\xi_1(0)\neq\xi_2(0)=0$. As in the real experiment, it is difficult to provide a stimulated emission for both modes of the twin beam due to different nonlinear processes which occur simultaneously. And as in the previous section, we will also denote the intensity of the initial coherent stimulating field $\xi_1(0)$ as $|\xi_1|^2$.
\subsubsection{Stimulated Pure Twin Beam on BS}
For a stimulated twin beam, the evolution matrices $ {\bf U} $ and $ {\bf V} $ in Eqs.~(\ref{UV}) and (\ref{CUV})  have
the following nonzero elements:
\begin{eqnarray}  
 U_{11}(t)&= &U_{22}(t)=\cosh(g_3t), \nonumber \\
 V_{12}(t)&=&V_{21}(t)=i\exp(i\alpha)\sinh(g_3t).
\end{eqnarray}
Without loss of generality, we can put the phase of the pump field $\alpha=0$.

The nonzero elements of the covariance matrix $\boldsymbol{ A_{\cal N}}$ are $B_1=B_2=B_{\rm p}$, where $B_{\rm p}=|V_{12}(t)|^2$ accounts for vacuum fluctuations of the stimulated twin beams, and it corresponds to the mean twin photon number in the case of the spontaneous emission, and $D_{12}=i\sqrt{B_{\rm p}(B_{\rm p}+1)}$. The mean photon number of the stimulated mode $k$ equals $\langle n_{k}\rangle = B_k + |\xi_k(t)|^2$, $k=1,2$. Note that $\xi_1(t)\neq\xi_2(t)$ as we assumed $\xi_2(0)=0$ in Eq.~(\ref{CUV}).

The beam splitter transformations on the stimulated twin beam state are applied in accordance with Eq.~(\ref{BSaction}).

\begin{figure}[t!] 
\includegraphics[width=0.35\textwidth]{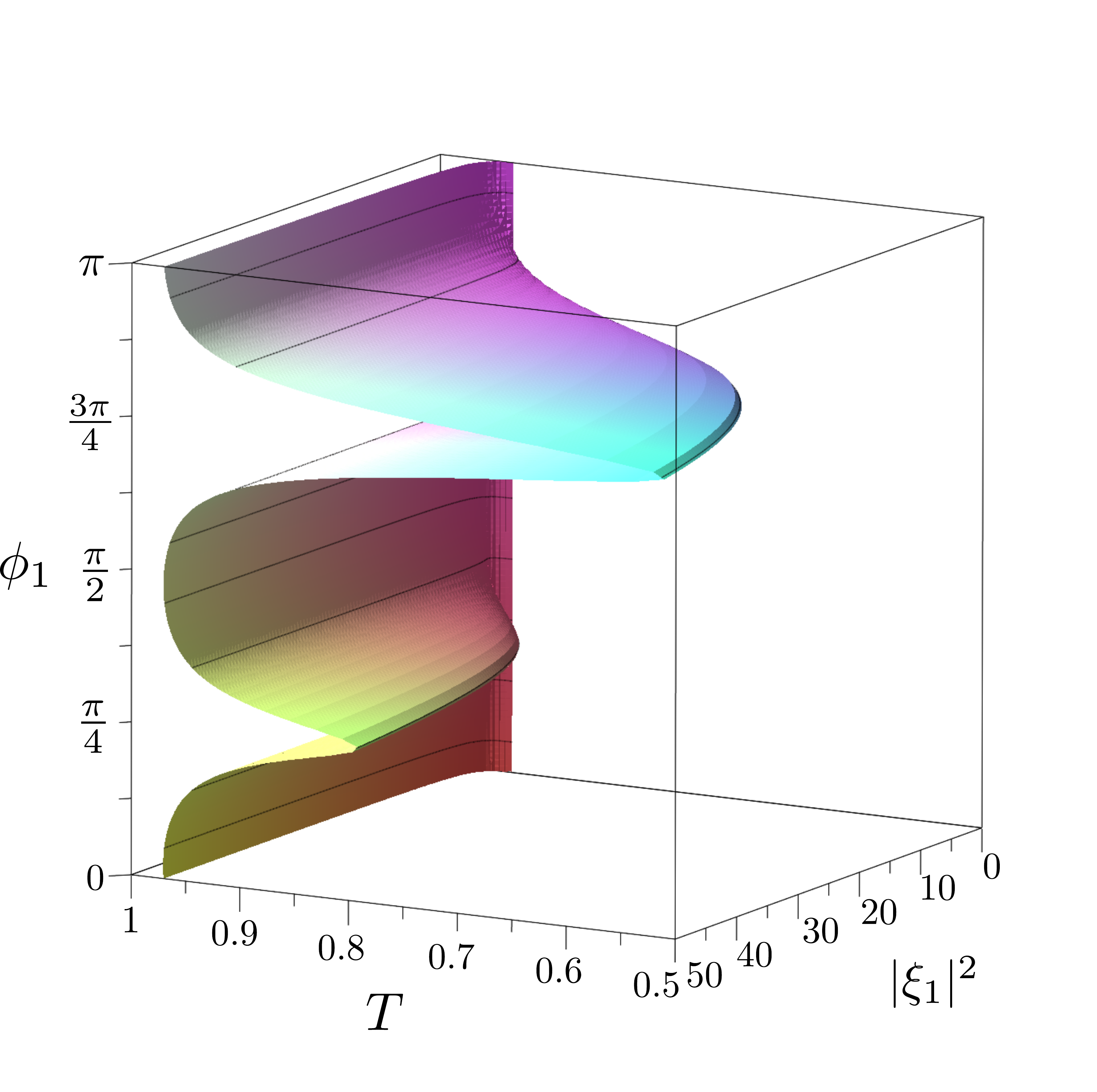}
\caption{Phase diagram of negativity of NW $M$ for a stimulated pure twin beam in space spanned by transmissivity $T$ of  BS,  intensity $|\xi_1|^2$, and phase $\phi_1$ of the stimulating coherent field, assuming that $B_{\rm p}=1$. The volume between the surface $M$ and plane $O|\xi_1|^2O\phi_1$ refers to the negativity of NW $M$. }\label{f4}
\end{figure}
\begin{figure} 
\includegraphics[width=0.35\textwidth]{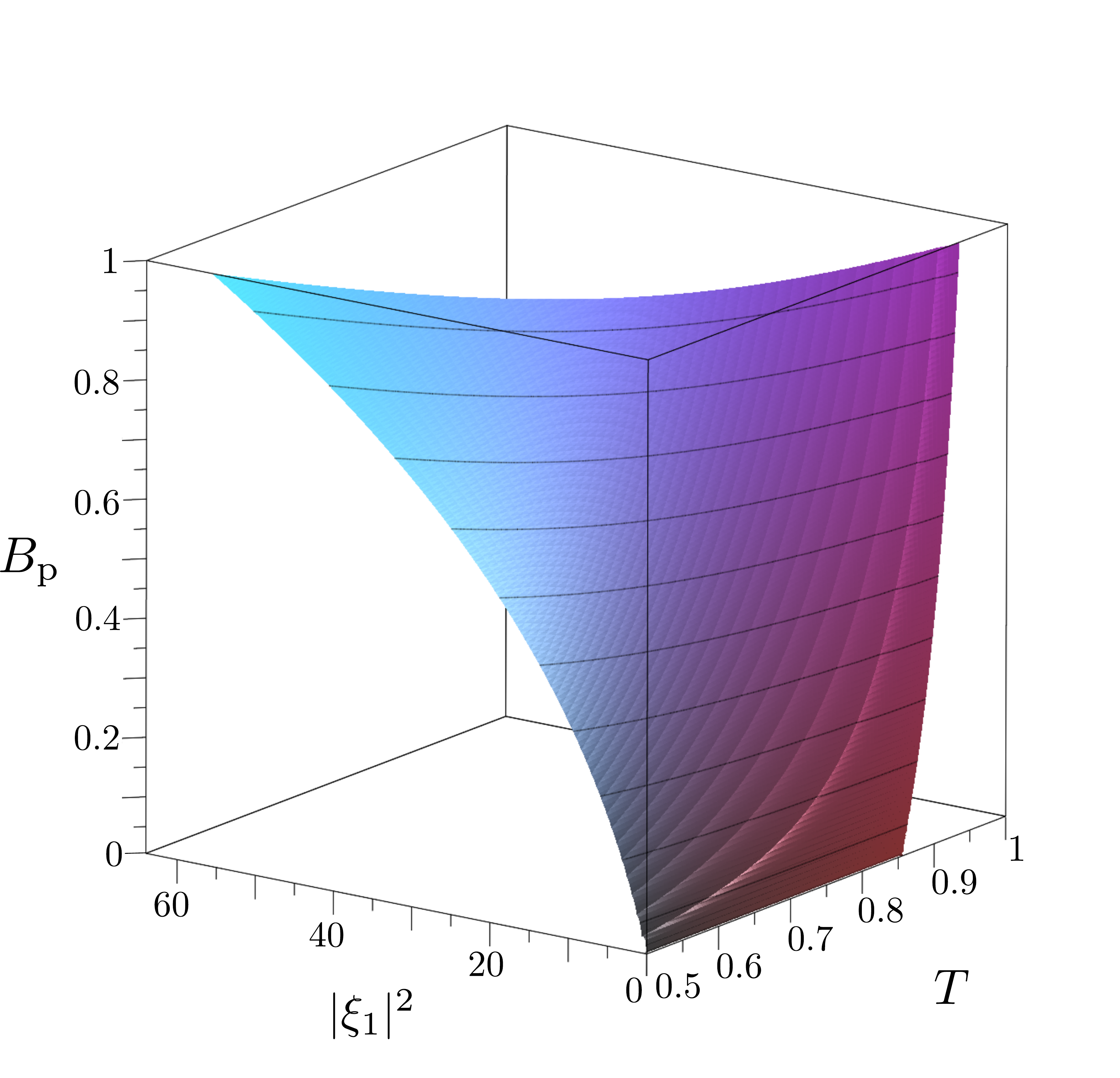}
\caption{Phase diagram of negativity of NW $M$ for a stimulated pure twin beam in space spanned by mean vacuum fluctuations $B_{\rm p}$ of the stimulated twin beam, transmissivity $T$ of  BS, and intensity $|\xi_1|^2$ of the stimulating coherent field, assuming that $\phi_1=3\pi/4+\pi n$. The volume below the surface refers to the negativity of NW $M$. }\label{f5}
\end{figure}
The greatest power of NW $M$ in the detection of the nonclassicality of a stimulated twin beam on BS with arbitrary $T$  is demonstrated when the phase $\phi_1$ of the stimulating field $\xi_1(0)=|\xi_1|e^{i\phi_1}$ equals $3\pi/4+\pi n$, $n\in{\mathbb Z}$ (see Fig.~\ref{f4}). One can also see that some power in the revealing of the nonclassicality by NW $M$ is displayed by values of $\phi_1=\pi/4+\pi n$, though it exhibits lower nonclassical sensitivity compared to $\phi_1=3\pi/4+\pi n$. By slowly varying $T:1\rightarrow1/2$, one needs to rely on larger intensities of stimulating field $|\xi_1|^2$ to reach the negative values of NW $M$. The same holds true when one increases the vacuum fluctuations $B_{\rm p}$ of the stimulated twin beam (see Fig.~\ref{f5}).

\begin{figure} 
\includegraphics[width=0.35\textwidth]{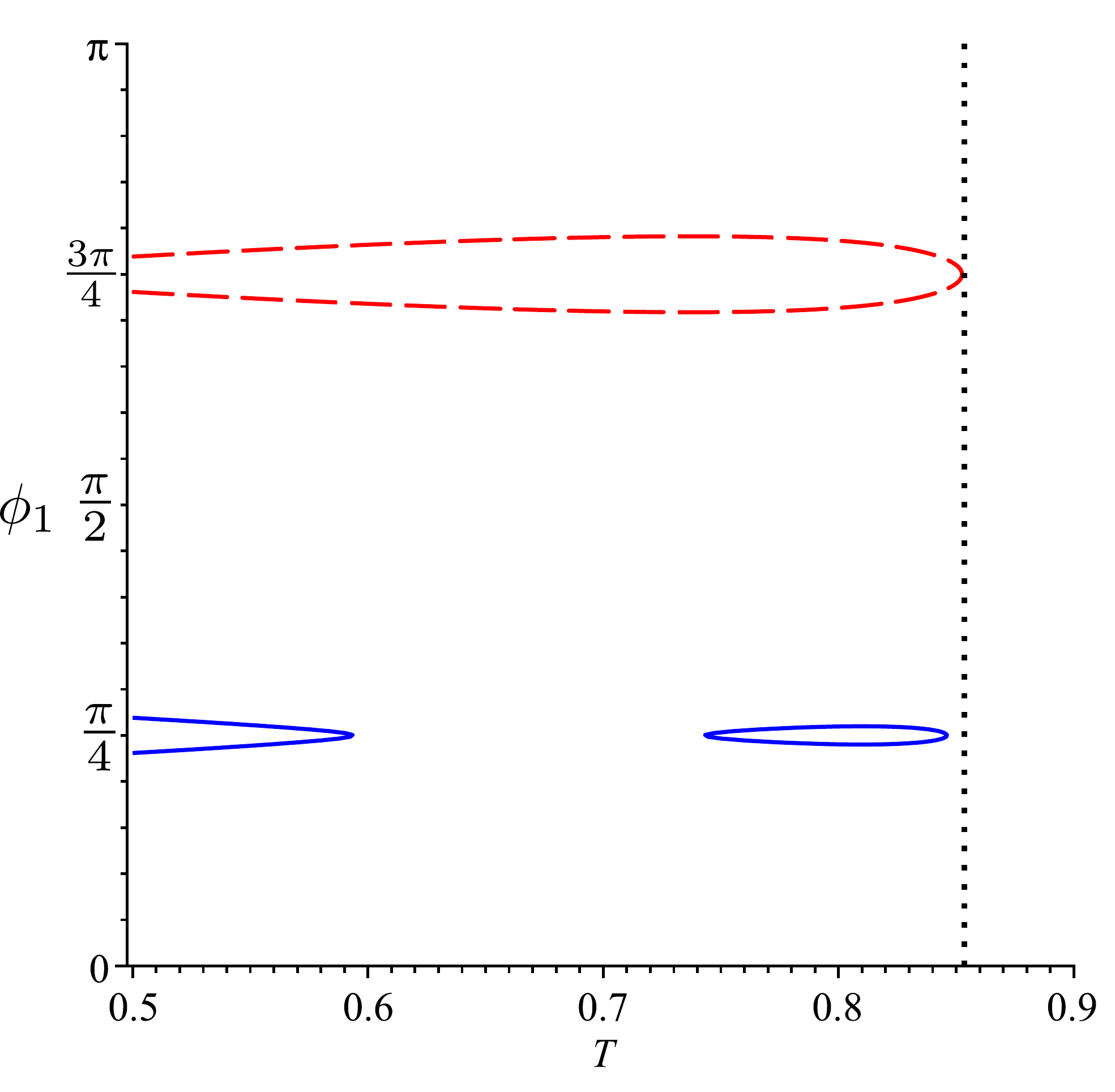}
\caption{Phase diagram of negative values of NWs $R_1$ (red dashed curve) and $R_{2}$ (blue solid curve) for stimulated pure twin beams in space spanned by transmissivity $T$ of BS and the phase $\phi_1$ of the stimulating coherent field, assuming $B_{\rm p}=1$, and $|\xi_1|^2=100$. The closed regions of each curve represent negative regions of the corresponding NW. For comparison LNI $I_{\rm ncl}^{(1)}=I_{\rm ncl}^{(1)}$ (black dotted curve) is shown on the graph which denotes the regions of the local nonclassicality (to the left from the black dotted  line). }\label{fig5}
\end{figure}

To detect the local nonclassicalities in both modes, one relies on NW $R_k$. The dependence of the negative values of $R_1$ and $R_2$ on the phase $\phi_1$ and transmissivity $T$ for a pure twin beam is presented in Fig.~\ref{fig5}. The graph indicates that the NWs $R_1$ and $R_2$ require different phases, namely, $\phi_1=3\pi/4$ and $\phi_1=\pi/4$, respectively, to extract the local nonclassicality of modes 1 and 2, correspondingly. Moreover, the nonclassicality detection power of $R_2$ is zero for some regions of $T$.  The latter fact is due to the constructive and destructive interference of the coherent components of the stimulated twin beam on the BS. Nonetheless, that problem can be solved by stimulating the second mode instead of the first one, i.e., $\xi_2(0)\neq\xi_1(0)=0$.
\subsubsection{Stimulated Twin Beam with Balanced Noise on BS}
For a stimulated twin beam which has the mean thermal noise photon number $B_{\rm s}$ ($B_{\rm i}$) in the signal (idler) mode, one has to modify only the quantity $B_1\rightarrow B_1=B_{\rm p}+B_{\rm s}$  ($B_2\rightarrow B_2=B_{\rm p}+B_{\rm i}$) in the initial covariance matrix $\boldsymbol{ A_{\cal N}}$ of the state before BS. In that case, we again adapt the model of the superposition of the quantum signal and noise, which we have already applied in the previous section. Additionally, in this section, we assume that the mean noise photon numbers in both arms of the stimulated twin beam are equal, $B_{\rm i}=B_{\rm s}$, i.e., the noise is balanced.
\begin{figure}[t!] 
\includegraphics[width=0.36\textwidth]{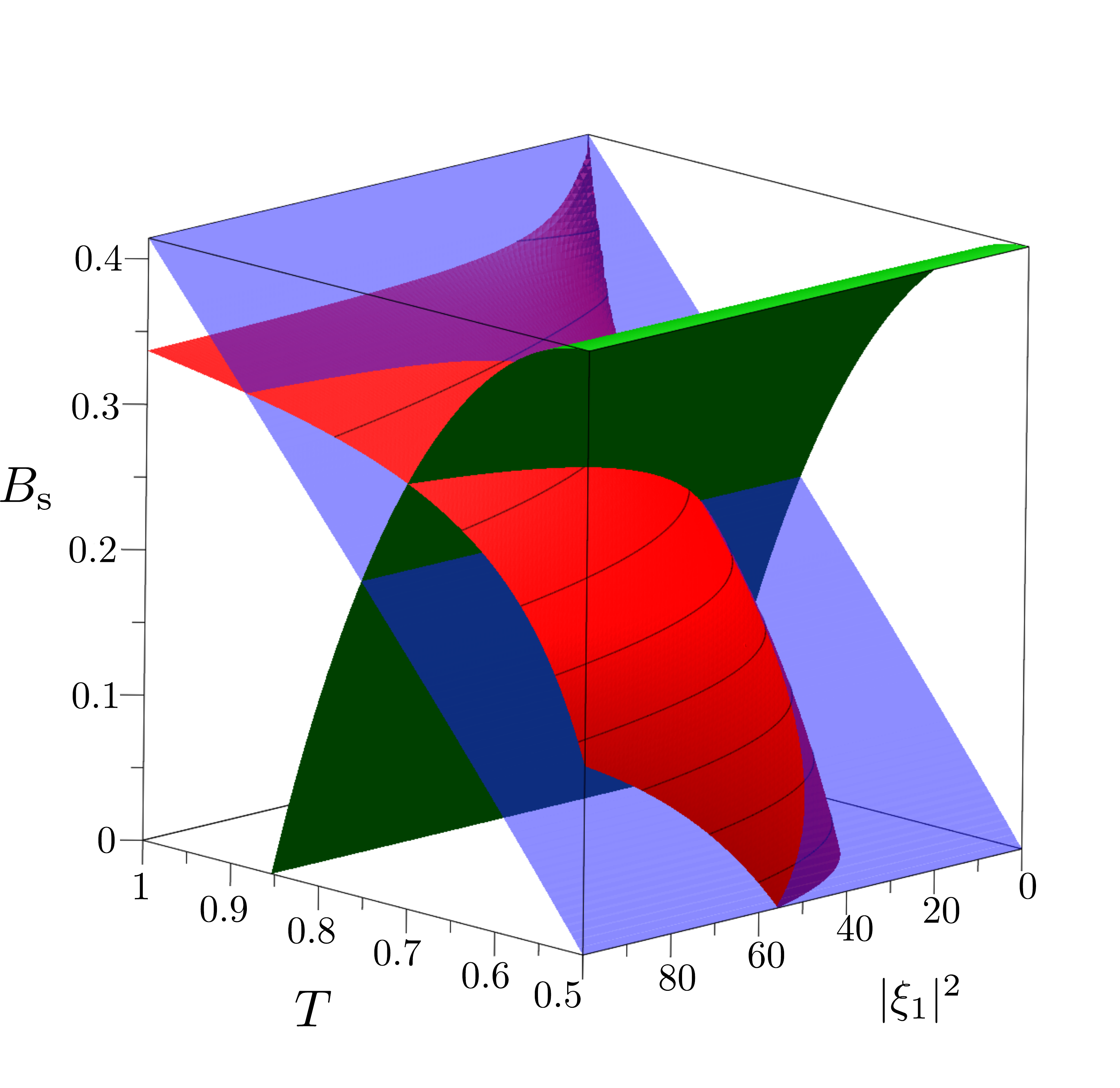}
\caption{Phase diagram of negativity of NW $M$ (red surface) for a stimulated noisy twin beam in space spanned by values of the mean noise photon number  $B_{\rm i}=B_{\rm s}$, transmissivity $T$ of the BS, and intensity $|\xi_1|^2$ of the stimulating coherent field, assuming $B_{\rm p}=1$, and $\phi_1=3\pi/4+\pi n$.  For comparison the nonclassicality diagrams of LNI $I_{\rm ncl}^{(1)}=I_{\rm ncl}^{(2)}$ (green surface), and EI $I_{\rm ent}$ (blue surface) are shown on the graph. The region below each surface refers to the corresponding negativity of $M$, local nonclassicality, and entanglement.}\label{fig4}
\end{figure}
\begin{figure} 
\includegraphics[width=0.35\textwidth]{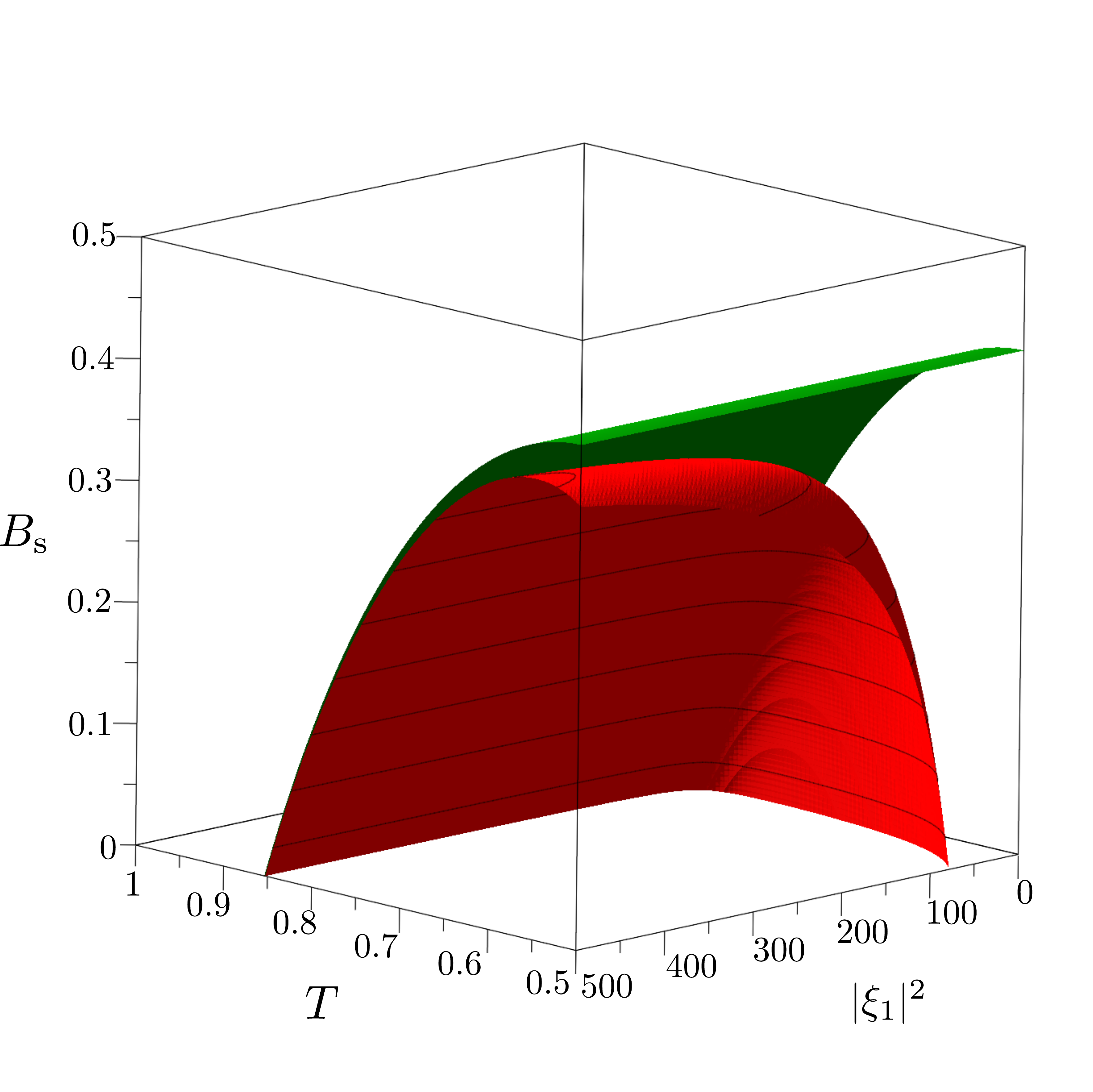}
\caption{Phase diagram of negativity of NWs $R_1$ (red surface) and LNI $I_{\rm ncl}^{(1)}$ (green surface) for a stimulated noisy twin beam in space spanned by mean noise photon number $B_{\rm i}=B_{\rm s}$, transmissivity $T$ of the BS, and intensity $|\xi_1|^2$ of the stimulating coherent field, assuming that $B_{\rm p}=1$, and $\phi_1=3\pi/4+\pi n$. The regions below each surface refer to the negative values of NW and local nonclassicality, respectively. }\label{fig6}
\end{figure}
As in the case of the stimulated pure twin beam, the optimal phase $\phi_1$  of the initial stimulating coherent field $\xi_1(0)$ should equal  $\phi_1=3\pi/4+\pi n$ in order to retrieve the local and global nonclassicality in terms of the given NWs for the noisy twin beam on BS.

 The typical behavior of the power of the NW $M$ for the stimulated noisy twin beam for the case when $B_{\rm p}=1$   is shown in Fig.~\ref{fig4}. One can see that the larger the noise $B_{\rm s}$ the larger intensities $|\xi_1|^2$ of the initial stimulating coherent field are required to access the nonclassicality. Moreover, even in case of the free-propagating twin beam $T=1$ for the noise $B_{\rm s}>1/3$, the NW $M$ fails to retrieve the nonclassicality (entanglement) regardless of the value of the intensity $|\xi_1|^2$. To overcome that difficulty, one can slightly perform coherent displacement of the second idler mode, which can be done, e.g., by stimulation emission of the idler beam. Nevertheless, even for a large range of the noise, the NW $M$ can identify the nonclassicality of the twin beam in the whole region of the transmissivity $T$. Most importantly, the NW $M$ along with entanglement is able to certify the local nonclassicality (see Fig.~\ref{fig4}). As such, one cannot explicitly distinguish whether the state is only entangled or squeezed by detecting the negativity of $M$ (see also Refs.~\cite{arkhipov2018a,arkhipov2018c}).

 Concerning the negativities of NW $R_1$, which identify the local squeezing,  Fig.~\ref{fig6} displays that the $R_1$ becomes a genuine NW of the local nonclassicality for any amount of noise in the system by increasing the intensity of the stimulating field $|\xi_1|^2$. The NW $R_2$, as it was pointed out before, in the case of the stimulated pure twin beam fails to become a genuine NW; nevertheless, one can add extra coherent displacement to the second idler mode to arrive at the negativities of NW $R_2$.
\subsubsection{Stimulated Twin Beam with Unbalanced Noise on BS}
\begin{figure} 
\includegraphics[width=0.36\textwidth]{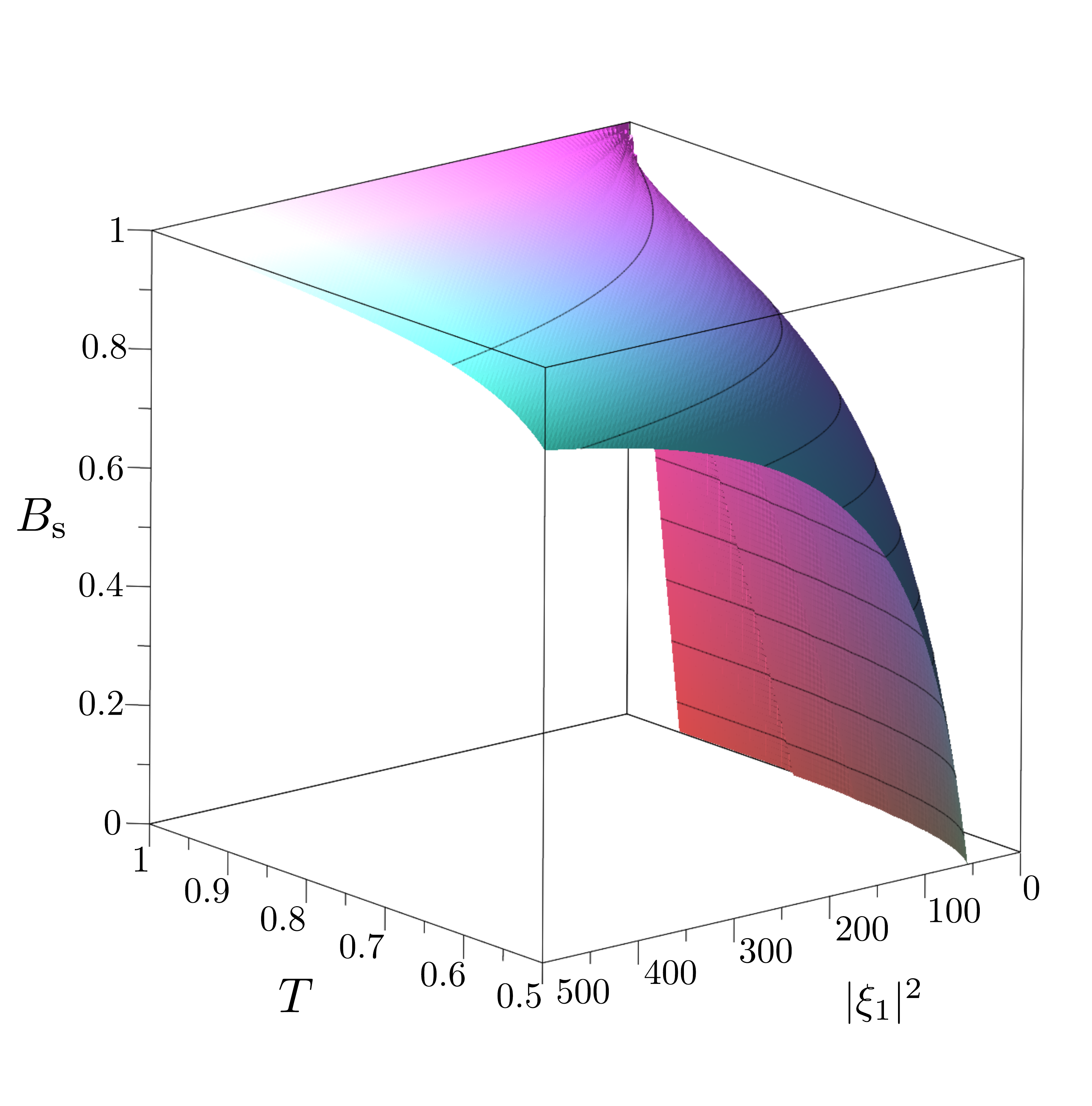}
\caption{Phase diagram of negativity of NW $M$  for a stimulated noisy twin beam in space spanned by  mean noise photon number  $B_{\rm s}$ ($B_{\rm i}=0$), transmissivity $T$ of the BS, and intensity $|\xi_1|^2$ of the stimulating coherent field, assuming that $B_{\rm p}=1$, and $\phi_1=3\pi/4+\pi n$. The regions below the surface refer to the region of  negative values of $M$. }\label{fig8}
\end{figure}
In the case of  asymmetrical noise present in two arms of the stimulated twin beam, e.g.,  when $B_{\rm i}=0$, the NW $M$ can completely reveal the nonclassicality of the stimulated noisy twin beam state (see Fig.~\ref{fig8}). Indeed, the nonclassicality of the twin beam state without stimulation on the BS should observe the nonclassicality for any $B_{s}<1$~\cite{arkhipov2015,arkhipov2016c}, though NW $M$ fails to detect it~\cite{arkhipov2018a}. But with induced stimulation emission of the given twin beam with unbalanced noise, the NW $M$ can unambiguously identify the nonclassicality of the stimulated twin beam state in the whole nonclassicality region spanned by the noise $B_{\rm s}$ and transmissivity $T$ for some value of the coherent intensity $|\xi_1|^2$ with initial phase $\phi_1=3\pi/4+\pi n$ (see Fig.~\ref{fig8}).

The NWs $R_1$ and $R_2$ have similar dependences on the noise $B_{\rm s}$ and transmissivity $T$ as in the case of the balanced noise.

\section{Multimode case}
In this section, we would like to briefly discuss the possibility to characterize the nonclassicality of the Gaussian states generated in the multimode subharmonic and down-conversion processes by means of induced stimulated emission. 

Since in the process of the multimode subharmonic generation the generated modes are separable and each mode is squeezed, one can straightforwardly apply the NW in Eq.~({13}) to each mode separately, provided that the appropriate phase of the stimulating coherent field is chosen for every mode. The same conclusion can be applied to the case of the $2n$-mode twin beam, where one deals with $n$ separable twin beams~\cite{arkhipov2015}. Therefore, by appropriately stimulating each twin beam and applying the NW given in Eq.~({14}), one can completely retrieve its pairwise multimode entanglement.


\section{Conclusions}
We have studied and compared the power of the nonclassicality witnesses given in Eqs.~(13) and (14)  in the detection of the local and global nonclassicality of the two-mode Gaussian states, generated in both the spontaneous and stimulated second subharmonic  and down-conversion processes, and which are subsequently subject to the beam splitter with arbitrary transmissivity. In the presented work, we have utilized the fact that the nonclassicality of the Gaussian states does not depend on the coherent part of the state which, in this study, has been expressed by a stimulating coherent field. Based on that fact, we have demonstrated that by means of the induced stimulated emission,  one can completely  identify the nonclassicality of the states previously generated in the spontaneous processes in terms of the given NWs.  As such, the induced stimulated emission of the parametric processes can be utilized in the complete experimental identification of the nonclassicality of the Gaussian states solely by means of the given NWs based on integrated intensity moments up to the third order. 

\acknowledgments  The author thanks Jan Pe\v{r}ina Jr. and Adam Miranowicz for fruitful and valuable discussions. This research was supported by  GA \v{C}R Project  No.~18-08874S.


%

\end{document}